\documentclass[aps,prb,twocolumn,superscriptaddress]{revtex4}

\usepackage{graphicx}
\usepackage{epstopdf}
\usepackage{hyperref}

\begin{document}


\title{Magnetic, electronic and transport properties of
high-pressure-synthesized chiral magnets Mn$_{1-x}$Rh$_{x}$Ge (B20)}


\author{V.~A.~Sidorov}
\affiliation{Vereshchagin Institute for High Pressure Physics, Russian Academy of Sciences,
108840 Troitsk, Moscow, Russia}

\author{A.~E.~Petrova}
\affiliation{Vereshchagin Institute for High Pressure Physics, Russian Academy of Sciences,
108840 Troitsk, Moscow, Russia}

\author{N.~M.~Chtchelkatchev}
\affiliation{Landau Institute for Theoretical Physics, Russian Academy of Sciences,
142432 Chernogolovka, Moscow Region, Russia}
\affiliation{Vereshchagin Institute for High Pressure Physics, Russian Academy of Sciences,
108840 Troitsk, Moscow, Russia}
\affiliation{Moscow Institute of Physics and Technology, 141700 Dolgoprudny, Moscow Region, Russia}

\author{M.~V.~Magnitskaya}
\affiliation{Vereshchagin Institute for High Pressure Physics, Russian Academy of Sciences,
108840 Troitsk, Moscow, Russia}
\affiliation{Lebedev Physical Institute, Russian Academy of Sciences, 119991 Moscow, Russia}

\author{L.~N.~Fomicheva}
\affiliation{Vereshchagin Institute for High Pressure Physics, Russian Academy of Sciences,
108840 Troitsk, Moscow, Russia}

\author{D.~A.~Salamatin}
\affiliation{Vereshchagin Institute for High Pressure Physics, Russian Academy of Sciences,
108840 Troitsk, Moscow, Russia}
\affiliation{Moscow Institute of Physics and Technology, 141700 Dolgoprudny, Moscow Region, Russia}

\author{A.~V.~Nikolaev}
\affiliation{Skobeltsyn Institute of Nuclear Physics, MSU, Vorob'evy Gory, 119991 Moscow, Russia}
\affiliation{Moscow Institute of Physics and Technology, 141700 Dolgoprudny, Moscow Region, Russia}

\author{I.~P.~Zibrov}
\affiliation{Vereshchagin Institute for High Pressure Physics, Russian Academy of Sciences,
108840 Troitsk, Moscow, Russia}

\author{F.~Wilhelm}
\affiliation{European Synchroton Radiation Facility, 38043 Cedex 9 Grenoble, France}

\author{A.~Rogalev}
\affiliation{European Synchroton Radiation Facility, 38043 Cedex 9 Grenoble, France}

\author{A.~V.~Tsvyashchenko}
\email{tsvyash@hppi.troitsk.ru}
\affiliation{Vereshchagin Institute for High Pressure Physics, Russian Academy of Sciences,
108840 Troitsk, Moscow, Russia}

\noaffiliation

\date{\today}

\begin{abstract}
We report on structural, magnetic and transport properties of a new set of the high-pressure-synthesized compounds Mn$_{1-x}$Rh$_x$Ge ($0 \leq x \leq 1$) with the chiral magnetic ordering. The magnetic and transport properties depend substantially on the concentration of rhodium ($x$) and the pressure. The saturation magnetic moment corresponds to a known high-spin value for pristine MnGe ($x = 0$) and decreases almost linearly with increasing concentration $x$. In addition, XMCD spectra taken at 10 K and 2 T indicate magnetic polarization of the Rh $4d$ electron states and Ge $4p$ states, which decreases with $x$, too. In rhodium rich compounds ($x \geq 0.5$) the temperature of the magnetic ordering increases significantly with pressure, whereas in manganese rich compounds ($x < 0.5$) the temperature decreases. Three different tendencies are also found for several structural and transport properties. In the intermediate range ($0.3 \leq x \leq 0.7$) samples are semiconducting in the paramagnetic phase, but become metallic in the magnetically ordered state. We carried out $ab~initio$ density-functional calculations of Mn$_{1-x}$Rh$_x$Ge at various concentrations $x$ and traced the evolution of electronic and magnetic properties. The calculation results are in good agreement with the measured magnetic moments and qualitatively explain the observed trends in transport properties.
\end{abstract}


\maketitle

\section{INTRODUCTION}
The discovery of bulk metal chiral magnets obtained by cobalt doping of a narrow-gap FeSi insulator with a very high anomalous Hall conductivity opens a new arena for spintronics \cite{1, 2} and prompts a search for new materials with such non-trivial properties. As another promising material we can mention a metastable high-pressure phase of MnGe, which as FeSi crystallizes in the B20 structure \cite{3}. The high pressure phase of MnGe displaying a large anomalous Hall effect \cite{4}, is a chiral magnet with the Neel temperature $T_{\mathrm{N}}$ = 170 K and the magnetic moment 2.3 $\mu_{\mathrm{B}}$ ($\mu_{\mathrm{B}}$ is the Bohr magneton) at 2 K \cite{5}. Unlike MnSi which demonstrates a quantum critical transition to a non-Fermi-liquid state under pressure \cite{6, 7}, the high-pressure phase of MnGe undergoes a two-stage transition. At the first stage (at 6 GPa) the magnetic moment of Mn transforms from a high spin state to a low spin state \cite{8}, while at the second stage at 23 GPa the magnetic moment disappears completing the transition to a paramagnetic phase \cite{9}.

Further research in this direction has become possible recently when on the basis of MnGe a new series of compounds - Mn$_{1-x}$Fe$_x$Ge with $x = [0.0; 1.0]$ has been synthesized under high pressure \cite{10}. It has been found that the Mn$_{1-x}$Fe$_x$Ge compounds crystallize in the same B20 structure and are magnetically ordered through the whole range of concentration. Small-angle neutron scattering reveals a ferromagnetic-like transition at $x_c = 0.75$ \cite{10}, where a short period helical structure characteristic of pure MnGe ($\left|k\right| = 2.3$ nm$^{-1}$) changes to a long period helical structure characteristic of pure FeGe ($\left|k\right| = 0.09$ nm$^{-1}$).

The magnetic structure found for Mn$_{1-x}$Fe$_x$Ge has been also observed in the Fe$_{1-x}$Co$_x$Ge series \cite{11}. In both these families the substitution involves solely $3d$-elements, and the increase of the number of electrons in the $3d$ band leads to a linear decrease of the cubic lattice constant (Vegard`s law) \cite{11}. This is not necessarily so if manganese is substituted with a $4d$ element, for example, with rhodium. The rhodium monogermanide - RhGe - also synthesized at high pressure, has the B20 crystal structure \cite{12} and displays unusual physical properties. It is a weak band ferromagnet with the ordering temperature $T_m = 140$ K, and at temperatures below $T_c = 4$ K the magnetic order in RhGe coexists with the superconductivity \cite{13}.

On the other hand, substituting Mn with Rh in MnGe with chiral magnetic ordering has resulted in the formation of magnetic spin helices with very long periods - ten times more than the period in the pristine compound \cite{14}. At the $x > 0.2$ level of the Rh-substitution, the neutron scattering shows patterns with a double periodicity. In Ref.~\cite{14} it has been suggested that the double periodicity reveals the presence of magnetic ``twist grain boundary'' phases, involving a dense short-range correlated network of magnetic screw dislocations.

In the present work we continue a comparative study of MnGe and RhGe and report on structural, magnetic and transport properties of a new set of the Mn$_{1-x}$Rh$_x$Ge compounds with the chiral magnetic ordering. Our main goal is the study of the evolution of physical characteristics when Mn is substituted with Rh. The paper is organized as follows. In Sec. II we give technical details of experimental techniques and theoretical calculations for these compounds. Experimental results are presented in Sec. III, followed by the density functional calculations and discussion, Sec. IV. In Sec. V we summaries the conclusions.

\section{EXPERIMENT AND THEORY}
Polycrystalline samples of Mn$_{1-x}$Rh$_x$Ge cubic phase were synthesized at 8 GPa and 1500 -1700 K using toroidal high-pressure apparatus \cite{15} by melting Mn, Rh and Ge \cite{16}. The purity was 99.9\% for Mn,  99.99\% for Rh and 99.999\% for Ge. The phase remains metastable after the pressure release. The crystal structure of samples was examined by x-ray diffraction (XRD). Measurements were performed at room temperature and ambient pressure using the diffractometer STOE IPDS-II (Mo-K$\alpha$) and Guinier camera e G670, Huber (Cu-K$\alpha_1$).

Magnetic properties were measured with VSM inserted in the PPMS. The electrical resistivity measurements were performed on bulk polycrystalline samples using a lock-in detection technique (SR830 lock-in amplifier) in the temperature range of 2 - 300 K. The magnetic ac-susceptibility was studied at pressures up to 6 GPa and temperatures down to 1.3 K with the use of a miniature clamped toroid-type device \cite{17}. Samples and pressure gauge (Pb) were placed in a Teflon capsule filled with liquid in this case.

Magnetic susceptibilities at high pressure were measured with the SR830 lock-in amplifier in home-made small coils placed inside teflon capsules filled with liquid. In all experiments the geometry of coils and the sample size were approximately the same. In this case a quantitative comparison between magnetic susceptibilities of different samples is possible though they are measured in arbitrary units (lock-in output voltage).

The x-ray absorption experiments were carried out at ID12 beamline at the European Synchrotron Radiation Facility (ESRF). The beamline is dedicated to polarization dependent x-ray-absorption studies\cite{18} in the energy range from 2 to 15 keV, that covers $K$-edges of Mn and Ge as well as $L_{2, 3}$ absorption edges of Rh. Sources of circularly polarized x-rays were helical undulators either of Helios-II type for experiments at Rh $L_{2, 3}$ or of Apple-II type for measurements at the $K$-edges of Mn and Ge. The spectra are recorded using total fluorescence yield detection mode in backscattering geometry. A magnetic field (up to 9 T) generated by a superconducting solenoid was applied along the x-ray beam direction and ensured a complete magnetic saturation of the samples at a temperature of 10 K. Magnetic saturation was additionally verified by recording so-called element-specific magnetization curves, i.e. intensity of XMCD at a given absorption edge as a function of applied magnetic field. To make sure that XMCD results are free of eventual experimental artifacts, the spectra were recorded by changing the helicity of the incoming x-rays for both parallel and antiparallel direction of the magnetic field.

Our $ab~initio$ computations are based on the density functional theory (DFT). We used the first-principles pseudopotential method as implemented in the Quantum Espresso package \cite{19}, with the exchange-correlation functional taken within generalized-gradient approximation (GGA) by Perdew-Burke-Ernzerhof (PBE) \cite{20}. We employed the projected-augmented-wave (PAW) type scalar-relativistic pseudopotentials from the SSSP library \cite{21}, with the valence electron configurations of $3s^2p^6d^54s^2$, $4s^2p^6d^75s^2$, and $3d^{10}4s^2p^2$ for Mn, Rh, and Ge, respectively. The integration over the irreducible Brillouin zone (BZ) for the electron density of states computation was performed on a uniform grid of $24 \times 24 \times 24$ $\mathbf{k}$-points. The B20 unit cell contains four formula units of Mn$_{1-x}$Rh$_x$Ge, that is four transition metal (TM) atoms and four metalloid atoms. For $x = 1/4, 1/2$, and 3/4, the due number of equivalent Mn atoms in the cell were replaced by Rh atoms. For $x = 1/8$ and 7/8 we used a $2 \times 1 \times 1$ supercell containing eight formula units. For $x = 15/16$, the $2 \times 2 \times 1$ supercell Mn$_{16}$Rh$_{15}$Ge$_{16}$ was used. The plane wave cutoff of 100 Ry was chosen, which gives the total energy convergence better than 10-8 Ry. For each composition, the equilibrium value of system`s lattice constant $a_0$ was defined as the one corresponding to zero pressure. The geometry relaxation was performed, until the residual atomic forces were converged down to 3 meV/\AA. The optimized internal atomic positions for MnGe are $u_{\mathrm{Mn}} = 0.135$ and $u_{\mathrm{Ge}} = 0.843$ (experimental values are 0.136 and 0.846), and for RhGe $u_{\mathrm{Rh}} = 0.122$ and $u_{\mathrm{Ge}} = 0.836$ (experimental values are 0.128 and 0.834). Fermi surface (FS) plots were generated using the XCrysDen software \cite{22}. Running ahead of discussion, we note that the structural and electronic properties of pure MnGe calculated here are in excellent agreement with those obtained in \cite{23} using the LAPW+lo method (Wien2k package) that is basically different from our approach.

\section{MEASUREMENT RESULTS}
\subsection{Crystal structure}

\begin{figure}
\includegraphics[width=0.5\textwidth]{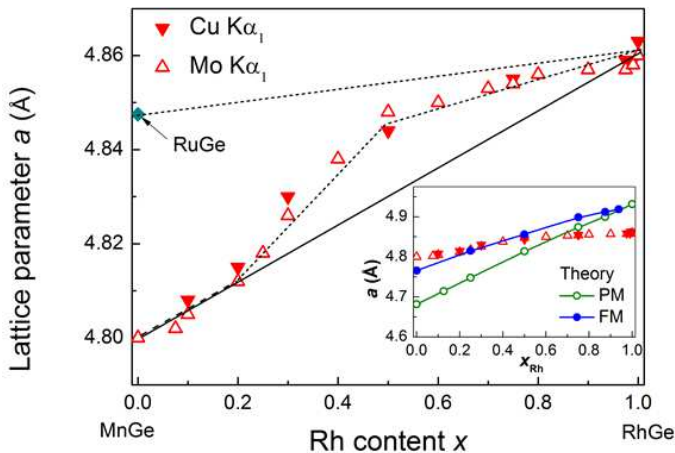}
\caption{\label{fig1}The dependence of lattice parameter a on the Rh content $x$ in Mn$_{1-x}$Rh$_x$Ge. Inset: Comparison with DFT-calculated $a(x)$ for the PM and FM states (see Section IV below).}
\end{figure}

We have found that in the whole range of concentrations $x$, the Mn$_{1-x}$Rh$_x$Ge compounds are crystallographically equivalent. All of them are crystallized in the same B20 cubic structure. In previous studies of the B20 systems Mn$_{1-x}$Fe$_x$Ge \cite{10}, Fe$_{1-x}$Co$_x$Ge \cite{11} and Mn$_{1-x}$Co$_x$Ge \cite{24, 25}, which are solid solutions of MnGe, FeGe and CoGe (monogermanides of 3d-metals), the linear change of the cubic lattice constant $a$ with $x$ has been observed (Vegard`s law). In contrast to these $3d$-solid solutions, Mn$_{1-x}$Rh$_x$Ge demonstrates an appreciable deviation from Vegard`s law, FIG.~\ref{fig1}. Starting with the lattice constant values for $x \approx 0$ and $x \approx 1$, one can draw the expected linear dependence for $a_{3d}(x)$, shown in FIG.~\ref{fig1} by black solid line. We see that the actual lattice constant function $a(x)$ follows Vegard`s law only up to $x \approx 0.25$. At concentrations $x > 0.5$, the dependence of $a$ on $x$ changes. Now it lies close to another (dotted black) line, $a_{4d}(x)$, which connects data for RuGe and RhGe, FIG.~\ref{fig1}. Clearly, this is the change from the 3d dependence to
the 4d one, since both Ru and Rh are 4d elements. Concentrations between $x = 0.25$ and $x = 0.5$ represent a transition region between 3d and 4d regimes.
FIG.~\ref{fig1} describes a peculiarity of the Mn$_{1-x}$Rh$_x$Ge system which holds for its other properties. It implies that properties of the manganese-rich samples ($x \approx 0$) differ from the rhodium-rich samples ($x \approx 1$).

\subsection{Magnetic properties}
\begin{figure}
\includegraphics[width=0.45\textwidth]{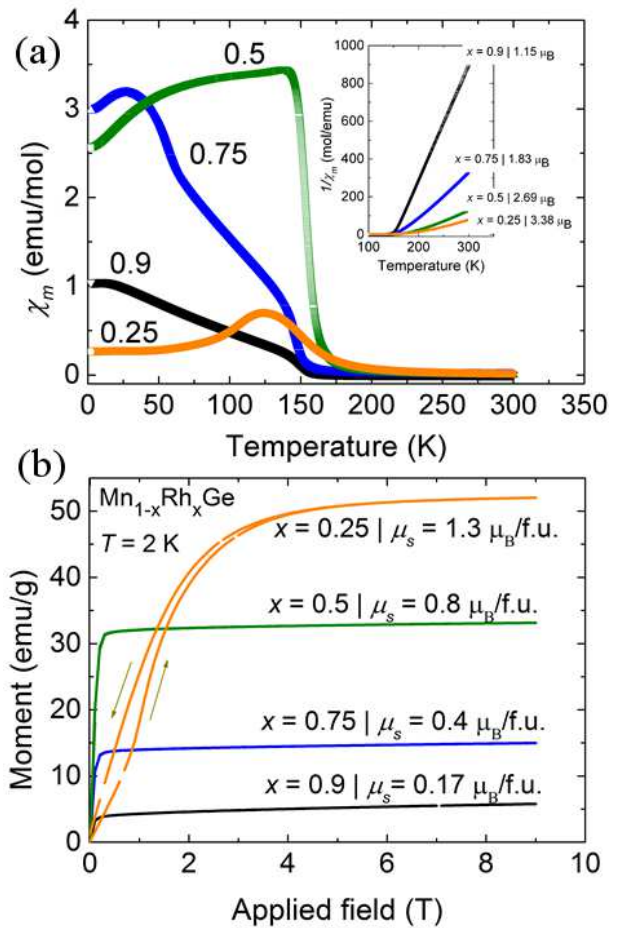}
\caption{\label{fig2}(a) Temperature dependences of magnetic susceptibility of Mn$_{1-x}$Rh$_x$Ge in the applied field $H = 300$ Oe. (b) Field dependences of magnetic moment of Mn$_{1-x}$Rh$_x$Ge at $T = 2$ K.}
\end{figure}

The temperature dependencies of the magnetic susceptibility ($\chi$) of Mn$_{1-x}$Rh$_x$Ge for some representative concentrations are presented in FIG.~\ref{fig2} (a). In the Mn-rich side ($x \leq 0.3$) the temperature dependence of susceptibility is similar to that of MnGe. With the increase of the rhodium concentration the broad peak signaling the onset of a chiral magnetic order is shifted down. For Mn$_{0.75}$Rh$_{0.25}$Ge this maximum is located at $T_{\mathrm{N}} = 125$ K whereas for pristine MnGe it is at $T_{\mathrm{N}} = 175$ K. At $x \geq 0.5$ the sharp step-like rise of the magnetic susceptibility at $T_m = 150-160$ K manifests the formation of the ferromagnetic-like order. FIG.~\ref{fig2} (b) shows the field dependence of the Mn$_{1-x}$Rh$_x$Ge magnetization. Notice, that between $x = 0.25$ and $x = 0.5$ there is a qualitative change of the plot shape. In the Mn-rich side (as in pristine MnGe) the field-induced ferromagnetic state appears only at high fields above 5 T \cite{4, 16}, whereas at $x \geq 0.5$ the magnetic moment saturates already at low fields 0.1-0.3 T.

\begin{figure}
\includegraphics[width=0.45\textwidth]{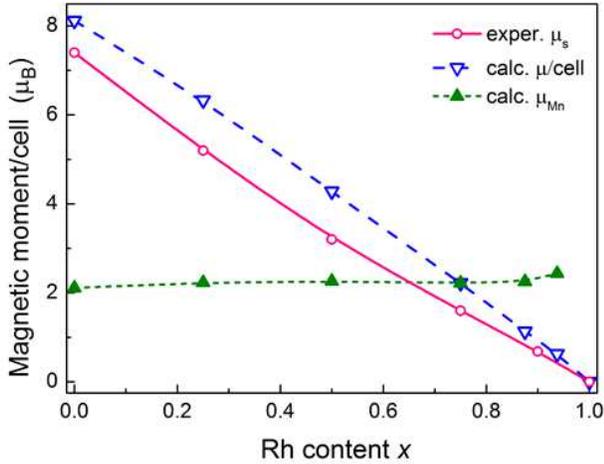}
\caption{\label{fig3}Magnetic moment of Mn$_{1-x}$Rh$_x$Ge as a function of Rh content $x$. The experimental $\mu_{\mathrm{s}}$ is shown by open circles. The down triangles denote the DFT-calculated spin moment per unit cell and the up triangles, per Mn atom (see Section IVB below). The lines are guide for the eye.}
\end{figure}

Though the magnetic properties of Mn$_{1-x}$Rh$_x$Ge are quite different for $x \leq 0.3$ and for $x \geq 0.5$, both the effective magnetic moment $\mu_{\mathrm{eff}}$ found by fitting $\chi^{-1}(T)$ (insert in FIG.~\ref{fig2} (a)) and the saturation moment $\mu_{\mathrm{s}}$ (FIG.~\ref{fig2} (b)) change nearly linearly through the whole range of rhodium concentrations, FIG.~\ref{fig3}.

\subsection{Transport properties}
\begin{figure}
\includegraphics[width=0.45\textwidth]{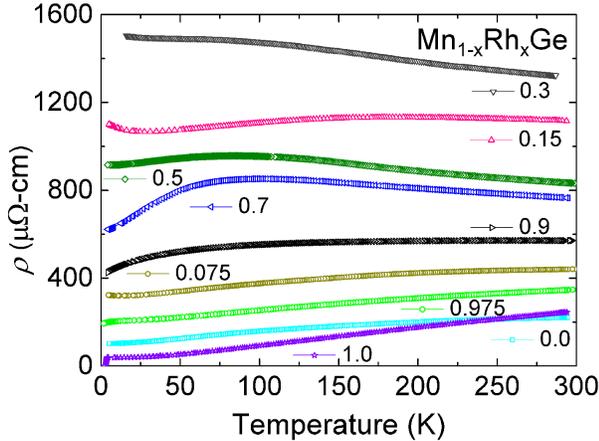}
\caption{\label{fig4}Temperature dependence of the electrical resistivity in Mn$_{1-x}$Rh$_x$Ge}
\end{figure}

The electric transport properties of Mn$_{1-x}$Rh$_x$Ge (FIG.~\ref{fig4}) exhibit unusual variation on going from $x = 0$ to $x = 1$. Both end members of a series MnGe and RhGe are metals with a moderate value of resistivity at room temperature $\approx 200$ $\mu \Omega$ cm. RhGe is a superconductor below $T_c \approx 4$ K. The addition of rhodium to MnGe produces the increase of resistivity of Mn$_{1-x}$Rh$_x$Ge and appearance of semiconducting behavior of $\rho(T)$ for $0.3 \leq x \leq 0.7$. The deviation from semiconducting behavior to metallicity for $x = 0.5$ and $x = 0.7$ seems to be correlated with the transition to the ferromagnetic state. It looks like Mn$_{1-x}$Rh$_x$Ge compounds at intermediate $x$ become semimetals or narrow gap semiconductors in the paramagnetic state. The transition to the magnetic state moves them to the metallic state. This unusual behavior of the electric transport properties indicates on the qualitative changes of the band structure of Mn$_{1-x}$Rh$_x$Ge in the paramagnetic and magnetically ordered state when rhodium concentration changes from $x = 0$ to $x = 1$.

\begin{figure}
\includegraphics[width=0.45\textwidth]{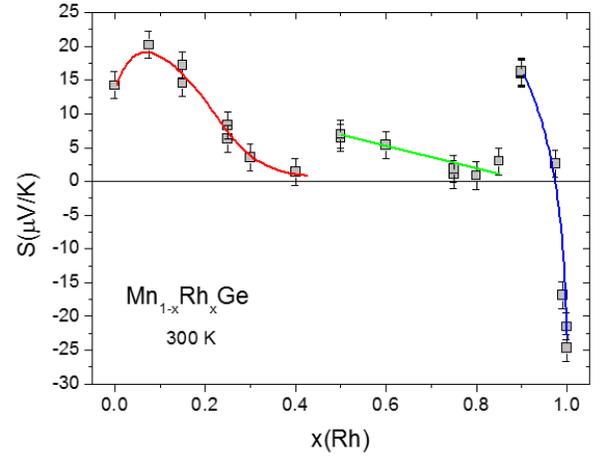}
\caption{\label{fig5}Concentration dependence of Seebeck coefficient for Mn$_{1-x}$Rh$_x$Ge measured at room temperature. Different points at a particular concentration $x$ refer to samples from different batches. Bold lines are guide for the eye.}
\end{figure}

The dependence of Seebeck coefficient $S$ on the rhodium concentration $x$ in Mn$_{1-x}$Rh$_x$Ge is shown in FIG.~\ref{fig5}. While MnGe has a positive value of $S = 15$ $\mu$V/K indicating the dominant role of hole carriers, RhGe has a negative value of $S = -25$ $\mu$V/K with electron carriers. The remarkable feature of this dependence is the presence of three distinct regions and the clear borders between them where Seebeck coefficient exhibits appreciable changes. With the increase of rhodium concentration from $x = 0$ to $x = 0.4$ the value of $S$ decreases and approaches zero at $x = 0.4$. But at $x = 0.5$ $S$ increases suddenly to the value of half of that of MnGe. Between $x = 0.5$ and $x = 0.8$ (the second region) $S$ decreases linearly and approaches zero again at $x = 0.8$. On further increase of $x$ from 0.8 to 0.9, the Seebeck coefficient increases sharply to 15 $\mu$V/K. The strongest changes of Seebeck coefficient are observed between $x = 0.9$ and $x = 1$ from +15 $\mu$V/K to -25 $\mu$V/K. It changes sign between $x = 0.975$ and $x = 0.99$. The anomalous behavior of $S$ at $x = 0.5$ and $x = 0.8$ is discussed further in Sec. IV where it is related with the electron density of states (DOS) at the Fermi level.

It is worth mentioning that a large negative Seebeck coefficient $S$ has been reported in non-magnetic CoGe \cite{26} and CoSi \cite{27}, which are isovalent with RhGe. Furthermore, $S(x)$ changes from a large negative to a large positive value at low hole doping when cobalt is substituted with iron (i.e. in Co$_{1-x}$Fe$_x$Ge and Co$_{1-x}$Fe$_x$Si). We observe the same behavior in Mn-doped RhGe (FIG.~\ref{fig5}).

\subsection{Pressure Effect}
\begin{figure*}
\includegraphics[width=\textwidth]{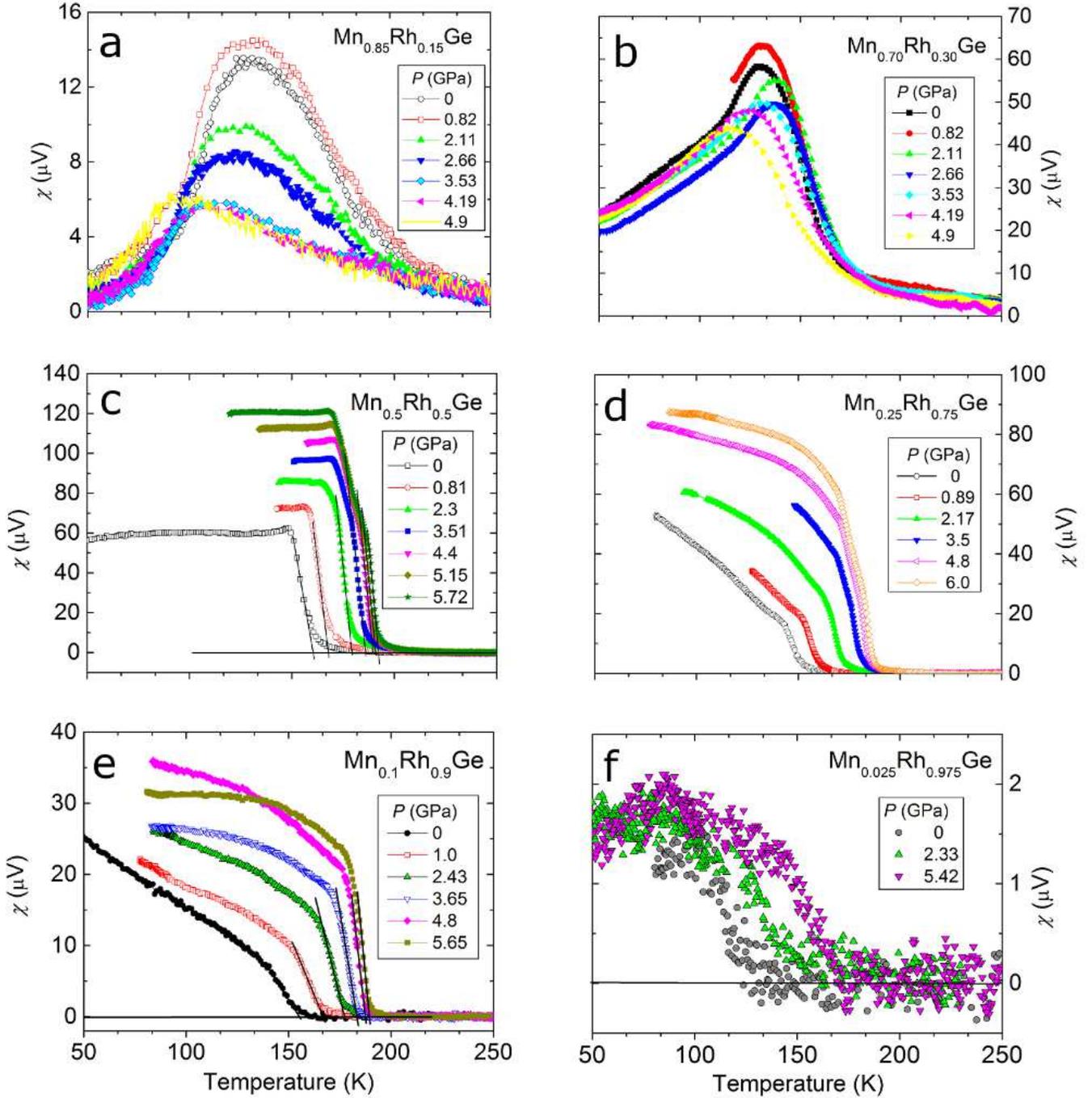}
\caption{\label{fig6}Temperature dependencies of the magnetic susceptibility of Mn$_{1-x}$Rh$_x$Ge at different pressures.}
\end{figure*}

Temperature dependences of the magnetic susceptibility ($\chi$) of Mn$_{1-x}$Rh$_x$Ge for various rhodium concentrations ($x = 0.15, 0.3, 0.5, 0.75, 0.9, 0.975$) at different pressures are presented in FIG.~\ref{fig6}a, b, c, d, e, correspondingly. Notice that the magnetic ordering temperature of Mn$_{1-x}$Rh$_x$Ge is very different for the regions $x < 0.5$ (FIG.~\ref{fig6}a, b, c) and $x \geq 0.5$ (FIG.~\ref{fig6}d, e, f).  In FIG.~\ref{fig6}a ($x = 0.15$) and FIG.~\ref{fig6} b ($x = 0.3$) we see that $\chi$ exhibits a broad peak at a certain (Neel) temperature $T_{\mathrm{N}}$. With increasing pressure the peak value ($\chi_{\mathrm{max}}$) decreases while $T_{\mathrm{N}}$ is shifted to lower temperatures. Earlier this behavior has been observed in MnGe \cite{8}. (In fact, at $x = 0.3$ where is a more complex pressure dependencies. The behavior discussed above holds only at high pressures [$P \geq 2.11$~GPa] while at low pressures [$P \leq 0.82$~GPa] the changes in $T_{\mathrm{N}}$ and $\chi_{\mathrm{max}}$ are reversed. We consider this as a precursor of the $x \geq 0.5$ tendency.)

The susceptibility behavior is completely changed in the range $x \geq 0.5$, FIG.~\ref{fig5}c, d, e, where $\chi(T)$ demonstrates a step-like temperature dependence at the transition temperature $T_m$. At $x = 0.5$ $\chi$ remains approximately constant at $T < T_m$, FIG.~\ref{fig6}c. At $x = 0.75$, $x = 0.9$ (FIG.~\ref{fig6}d and \ref{fig6}e) $\chi$ in the region $T < T_m$ increases with decreasing temperature. The character of pressure dependency of $\chi$ at $x \geq 0.5$ is also changed: with increasing pressure $T_m$ is shifted up while $\chi_{\mathrm{max}}$ increases.

\begin{figure}
\includegraphics[width=0.45\textwidth]{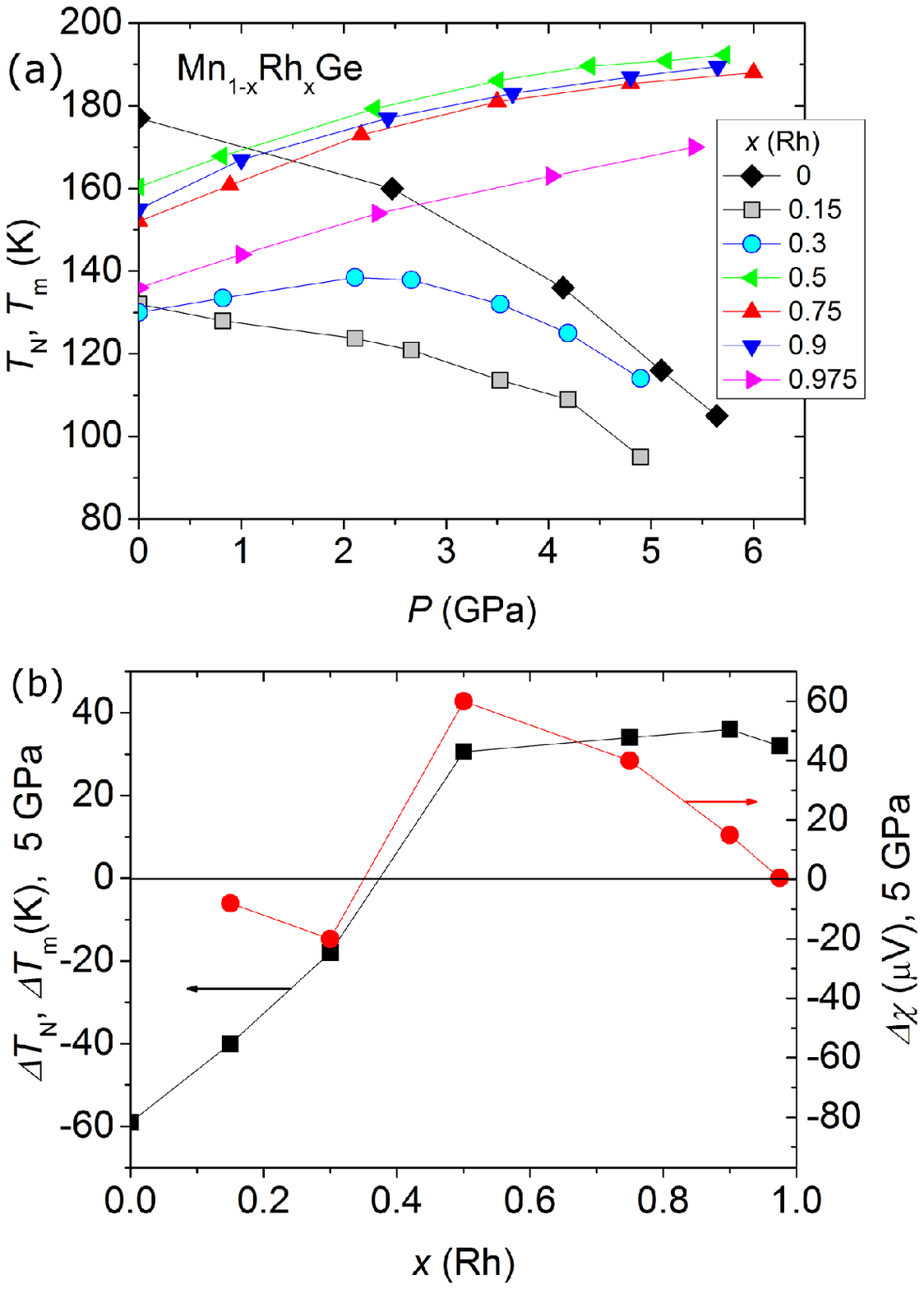}
\caption{\label{fig7}(a) Pressure dependencies of the magnetic ordering temperatures $T_{\mathrm{N}}$, $T_m$ for different compounds in the system Mn$_{1-x}$Rh$_x$Ge (b) Black squares - shift of the magnetic ordering temperature $T_{\mathrm{N}}$, $T_m$ for different compounds in the system Mn$_{1-x}$Rh$_x$Ge at 5 GPa. Red circles - changes of the amplitude of susceptibility peak (step) at the magnetic transition at 5 GPa compared to that at ambient pressure.}
\end{figure}

Two different types of pressure dependencies of Mn$_{1-x}$Rh$_x$Ge are clearly illustrated in FIG.~\ref{fig7}a and \ref{fig7}b. FIG.~\ref{fig7}a shows the pressure dependencies of $T_{\mathrm{N}}$ for concentrations $x \leq 0.3$ (type I) and $T_m$ for $x \geq 0.5$ (type II). FIG.~\ref{fig7}b shows the rhodium concentration dependence of the quantity $\Delta T_{\mathrm{N}} = T_{\mathrm{N}}(0) - T_{\mathrm{N}}(P_1)$ or $\Delta T_m = T_m(0) - T_m(P_1)$ [black line with black squares] and the quantity $\Delta \chi = \chi_{\mathrm{max}}(0) - \chi_{\mathrm{max}}(P_1)$ [red line and red circles], where $P_1 = 5$~GPa. Thus, $\Delta T_{\mathrm{N}}$ [or $\Delta T_m$] and $\Delta \chi$ signify a change of these pressure derived quantities with $x$. From FIG.~\ref{fig7}a we clearly see two types of dependencies: decreasing temperature values of $T_{\mathrm{N}}$ for $x \leq 0.3$ and monotonically increasing values of $T_m$ for $x \geq 0.5$. FIG.~\ref{fig7}b also demonstrates the two types of tendencies, the first region with negative values of $\Delta T_{\mathrm{N}}$ and $\Delta \chi$ at $x \leq 0.3$ and the second with the positive values of $\Delta T_m$ and $\Delta \chi$ at $x \geq 0.5$.

\subsection{XANES and XMCD measurements}
\begin{figure}
\includegraphics[width=0.45\textwidth]{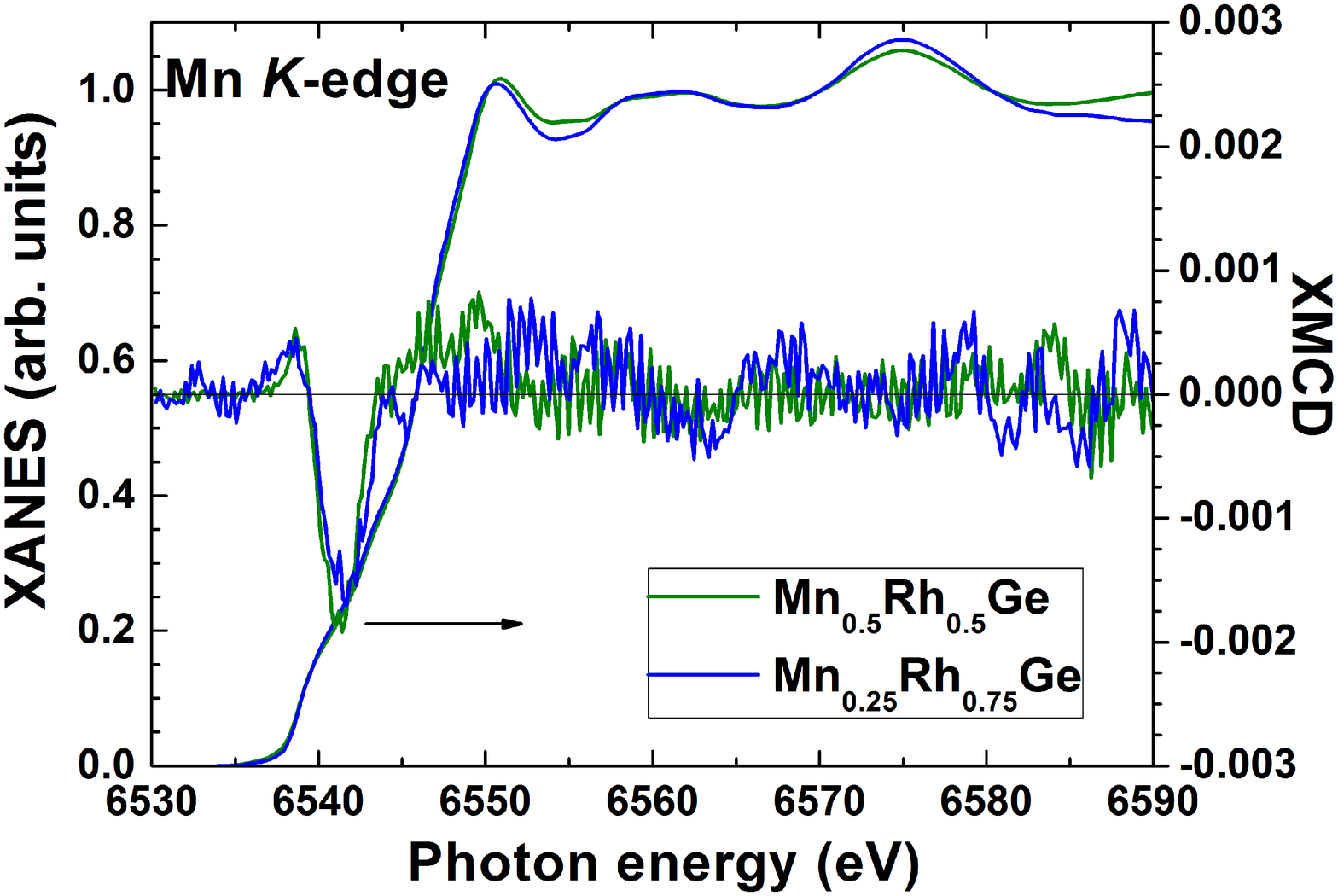}
\caption{\label{fig8}Mn $K$-edge XANES and XMCD spectra, measured at 10 K and 2 T.}
\end{figure}

\begin{figure}
\includegraphics[width=0.45\textwidth]{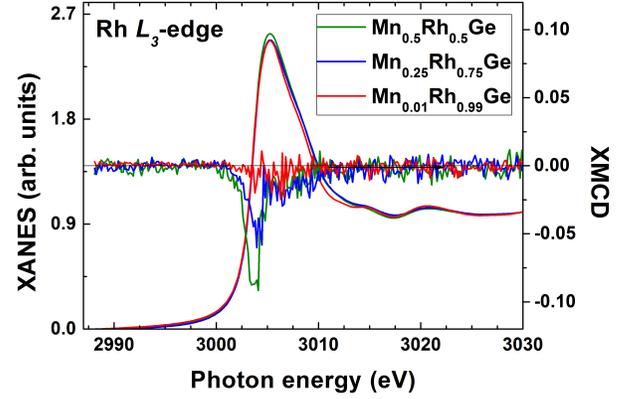}
\caption{\label{fig9}Rh $L_3$-edge XANES and XMCD spectra, measured at 10 K and 2 T. Mn$_{0.01}$Rh$_{0.99}$Ge measured at 2 K and 9 T.}
\end{figure}

\begin{figure}
\includegraphics[width=0.45\textwidth]{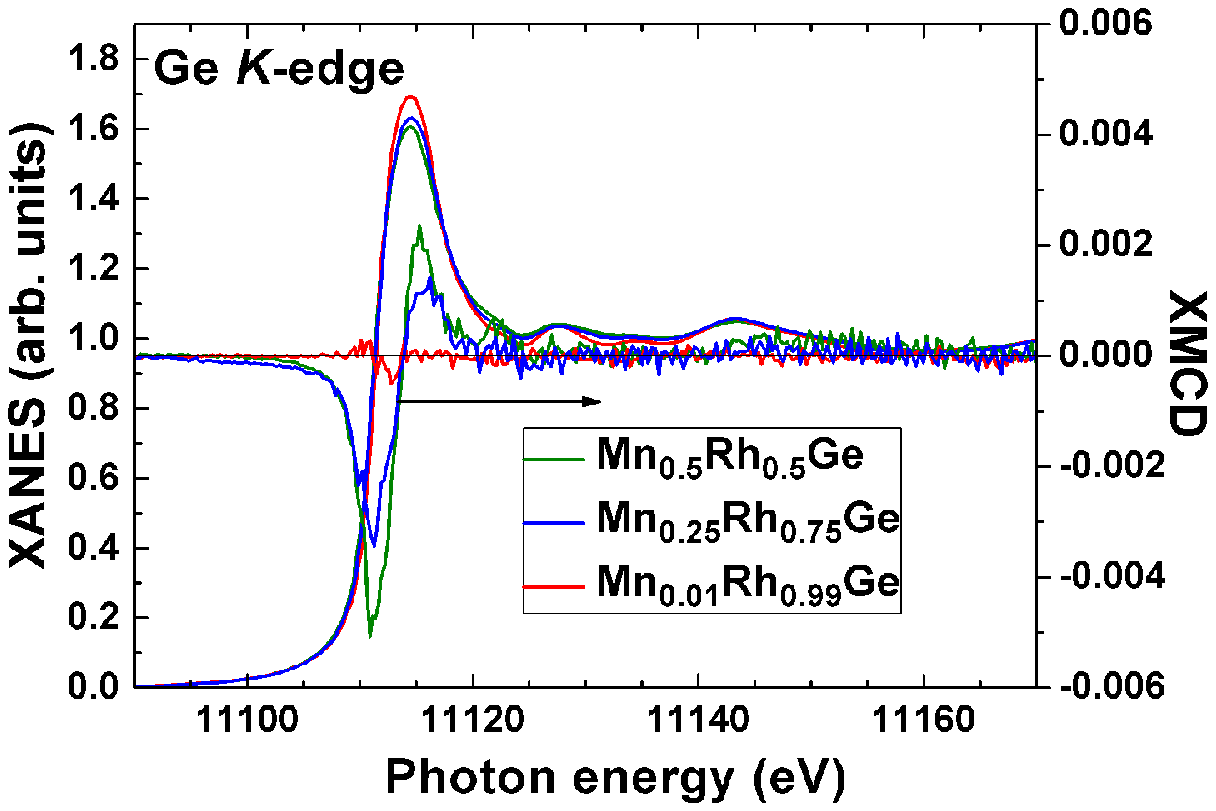}
\caption{\label{fig10}Ge $K$-edge XANES and XMCD spectra, measured at 10 K and 2 T. Inset shows the XMCD spectrum of the Ge $K$-edge for Mn$_{0.01}$Rh$_{0.99}$Ge measured at 2 K and 9 T.}
\end{figure}

Experimental XANES and XMCD spectra at the $K$-edge of Mn, $L_3$-edge of Rh and $K$-edge of Ge are presented in FIGs.~\ref{fig8}, \ref{fig9} and \ref{fig10}, respectively. The spectra have been corrected for self-absorption effects and for the incomplete circular polarization rate: 0.05 at the $L_3$ edge of Rh, 0.85 at the Mn $K$-edge and 0.95 at the $K$-edge of Ge. The XANES and XMCD spectra provide information on the electronic and magnetic properties of Mn, Rh and Ge atoms in Mn$_{1-x}$Rh$_x$Ge.

One can see on FIG.~\ref{fig8} that the XANES spectra recorded at the Mn $K$-edge for samples with $x=0.5$ and $x=0.75$ have a very similar spectral shape. This indicates that the valence state of the Mn atoms changes barely if at all between these two samples. Unfortunately, the XANES spectrum could not be recorded for the sample with $x=0.99$ due to a very strong fluorescence and x-ray scattering background signal. The samples with $x=0.5$ and $x=0.75$ exhibit sizeable XMCD signals at the $K$-edge of Mn under 2 T field and at temperature of 10 K. The normalized amplitude and the spectral shape of XMCD signals shown on FIG.~\ref{fig8} are also similar for both samples. It implies that the magnetic moment carried by Mn atoms remains nearly the same for both samples. This
observation is in good agreement with the results of present DFT-calculations. XANES and XMCD spectra recorded at the $L_3$-edge of Rh for all three samples are reproduced on FIG.~\ref{fig9}. The XANES spectra for all three samples look rather similar. It shows that there are no significant changes in the occupation of $4d$ states of Rh with concentration, even though some minor increase in number of $4d$ holes is present in the sample with $x=0.5$. The negative sign of the XMCD signal at the $L_3$ absorption edge indicate that the induced magnetic moment of Rh is parallel to an applied magnetic field and, therefore, to the magnetic moment of Mn~\cite{29}. Unfortunately, the presence of a strong diffraction peak around $\approx 3.16$ keV prevents us from measuring ``clean'' XANES and XMCD spectra at the $L_2$-edge of Rh. Consequently, we could not use the sum rules analysis to determine the spin and orbital magnetic moments induced in $4d$ shells of Rh atoms. However, the intensity of XMCD signals at the $L_3$-edge of Rh for all three samples follows the experimentally measured values of macroscopic magnetization. To get further insight into magnetism of Rh in the studied samples we exploit XMCD recorded on a reference system for the induced magnetic moment on Rh, namely FeRh films \cite{30Rogalev}. Comparing the present XMCD signals with the reference one, we could estimate the induced magnetic moment carried by Rh atoms to be about +0.1~$\mu_{\mathrm{B}}$ in Mn$_{0.5}$Rh$_{0.5}$Ge, +0.06$\mu_{\mathrm{B}}$ in the sample with $x=0.75$ whereas it is below 0.01 $\mu_{\mathrm{B}}$, if any, in Mn$_{0.01}$Rh$_{0.99}$Ge sample. The XANES and XMCD spectra measured at the Ge $K$-edge for all three samples are shown on FIG.~\ref{fig10}. The XANES spectra show some changes in the white-line intensity at the Ge $K$-edge, which could be attributed to a minor changes in the occupation of the $4p$ states of Ge with increase of concentration of Mn. The XMCD spectra for $x=0.5$ and 0.75 samples display a dispersive like shape with a negative peak at 11.11 keV. Amplitude of XMCD signals at the Ge $K$-edge is comparable to those at the $K$-edge of Mn and that implies a significant magnetic polarization of the $4p$ states of germanium, which scales rather well with the macroscopic magnetization. In Mn$_{0.01}$Rh$_{0.99}$Ge with a very weak total magnetization, the XMCD signal is at the detection limit. Thus, XMCD results show that induced magnetic moments on both Rh and Ge atoms are closely related to the exchange field due to Mn magnetic moments.

\section{CALCULATION RESULTS AND DISCUSSION}
We calculated in detail the electronic properties of the system Mn$_{1-x}$Rh$_x$Ge for $x$ changing from 0 to 1, with and without taking account of the spin polarization. The experimentally measured helix pitch is very large in comparison with lattice parameter, therefore our spin-polarized calculations were done using a simple model of collinear ferromagnetism. Both nonmagnetic and magnetic solutions are obtained for all the compounds, except RhGe that turned out to be nonmagnetic (see below). The cubic lattice parameter a of the system increases with increasing Rh content, as shown in the inset to FIG.~\ref{fig1}. This could be expected, because the atomic radius of rhodium is a few percent larger than that of manganese. For the both PM and FM states, the concentration dependence of lattice parameter, $a(x)$, is practically linear, in accordance with Vegard`s law. Thus, the anomalous behavior of $a(x)$ observed experimentally at room temperature (see FIG.~\ref{fig1}) is not predicted in our DFT calculations. Note that the dependence $a(x)$ for Mn$_{1-x}$Rh$_x$Ge in the FM state has been previously calculated at $0 \leq x \leq 0.5$ \cite{14}.

The theoretical lattice parameter a of nonmagnetic RhGe (4.931~\AA) exceeds the experimental one by ~1.5\%, which is acceptable accuracy within DFT-GGA approach. Our results for pure RhGe partially published in Ref.~\cite{30} are consistent with the theoretical data \cite{13}. For the compounds Mn$_{1-x}$Rh$_x$Ge in the FM state, a disagreement between the theoretical and experimental curves a(x) is within uncertainty of measurements and calculations (for MnGe, $a_{\mathrm{FM}} = 4.766$~\AA~is only 0.5\% less than the measured value of 4.79~\AA~\cite{31}). The same is true for the PM state at $x > 0.25$. At the same time, the calculated a value is strongly underestimated for the Mn-rich side of the series Mn$_{1-x}$Rh$_x$Ge. The lattice parameter of PM MnGe ($a_{\mathrm{PM}} = 4.682$~\AA) is by ~2.6\% smaller than the room-temperature value (~2.2\% as compared with a value extrapolated to low temperatures using the experimental thermal expansion \cite{31}. Similar results for pure MnGe in the PM and FM states have been obtained in the DFT calculations \cite{23, 32}. Such a deviation is understandable, because correlation effects are important in systems with partially filled $3d$ bands, especially in the middle of the $3d$ period, moreover, these effects are stronger for Mn, than for other $3d$-metals \cite{33}. It is also well known that the DFT-calculated atomic volume and bulk modulus of manganese (and some Mn-based compounds) are respectively, underestimated and overestimated in non-spin-polarized calculations, but become more reasonable if spin polarization is taken into account \cite{34}. At the same time, such calculations of $4d$ systems provide rather good agreement with experiment. Reliable values of a on the Mn-rich side of Mn$_{1-x}$Rh$_x$Ge in the PM state can be obtained within `beyond-DFT` approaches (such as DMFT) not used here.

The theoretical bulk modulus $B$ of MnGe is known to be of 133 GPa, which exceeds the available experimental value of 106 GPa (see Ref.~\cite{8} and references therein). Our approximate estimate gives $B = 135$ GPa for RhGe and 154 GPa for Mn$_{0.5}$Rh$_{0.5}$Ge.

\subsection{Non-spin-polarized calculations}
\begin{figure}
\includegraphics[width=0.45\textwidth]{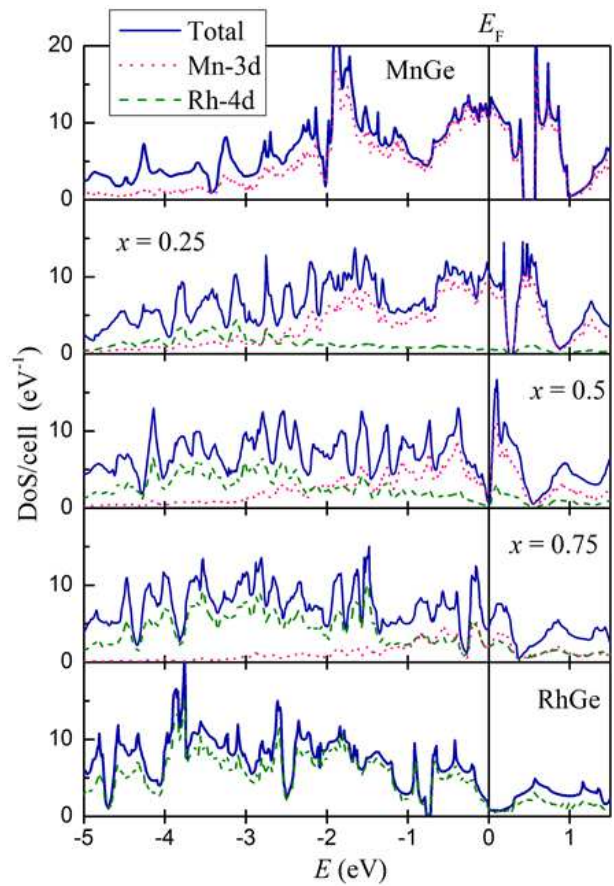}
\caption{\label{fig11}From top to bottom: The evolution of the total, $N(E)$, and $3d$-, $4d$-projected density of states in paramagnetic Mn$_{1-x}$Rh$_x$Ge with increasing $x$. The Rh content $x$ is indicated on corresponding panels (a)-(e). The Fermi energy $E_{\mathrm{F}}$ is at 0 eV.}
\end{figure}

\begin{figure}
\includegraphics[width=0.5\textwidth]{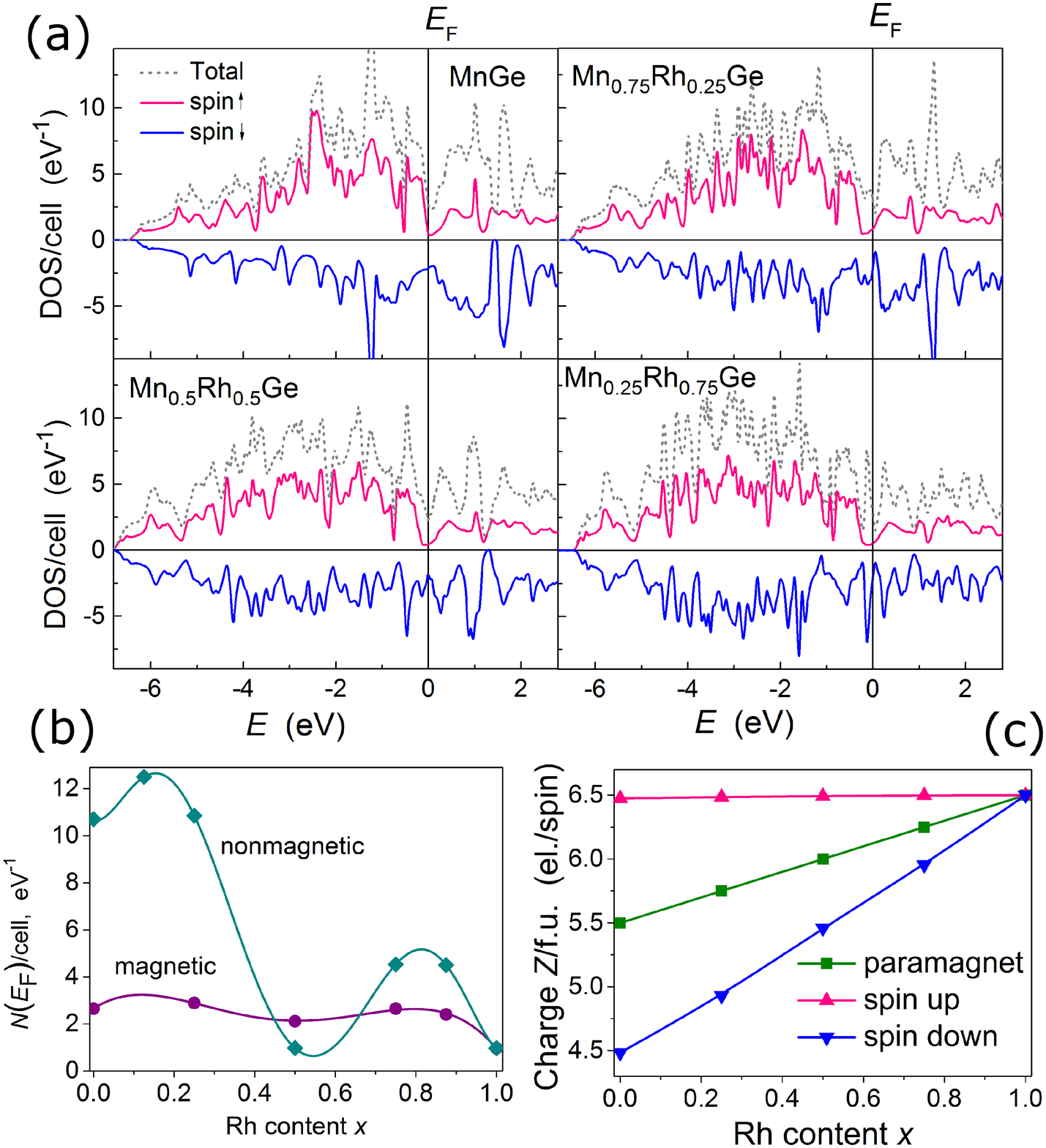}
\caption{\label{fig12}(a) The total, spin-up and spin-down DOS $N(E)$ at $x = 0, 0.25, 0.5$ and 0.75. The Fermi energy $E_{\mathrm{F}}$ is at 0 eV, and $N(E)$ for the spin up (down) states is counted positive (negative). In each case, the spin-up DOS is rather similar to that of nonmagnetic RhGe in FIG.~\ref{fig11}e. (b), (c) The DOS at $E_{\mathrm{F}}$ and valence $Z$, respectively, in magnetic and paramagnetic case. }
\end{figure}

\begin{figure}
\includegraphics[width=0.5\textwidth]{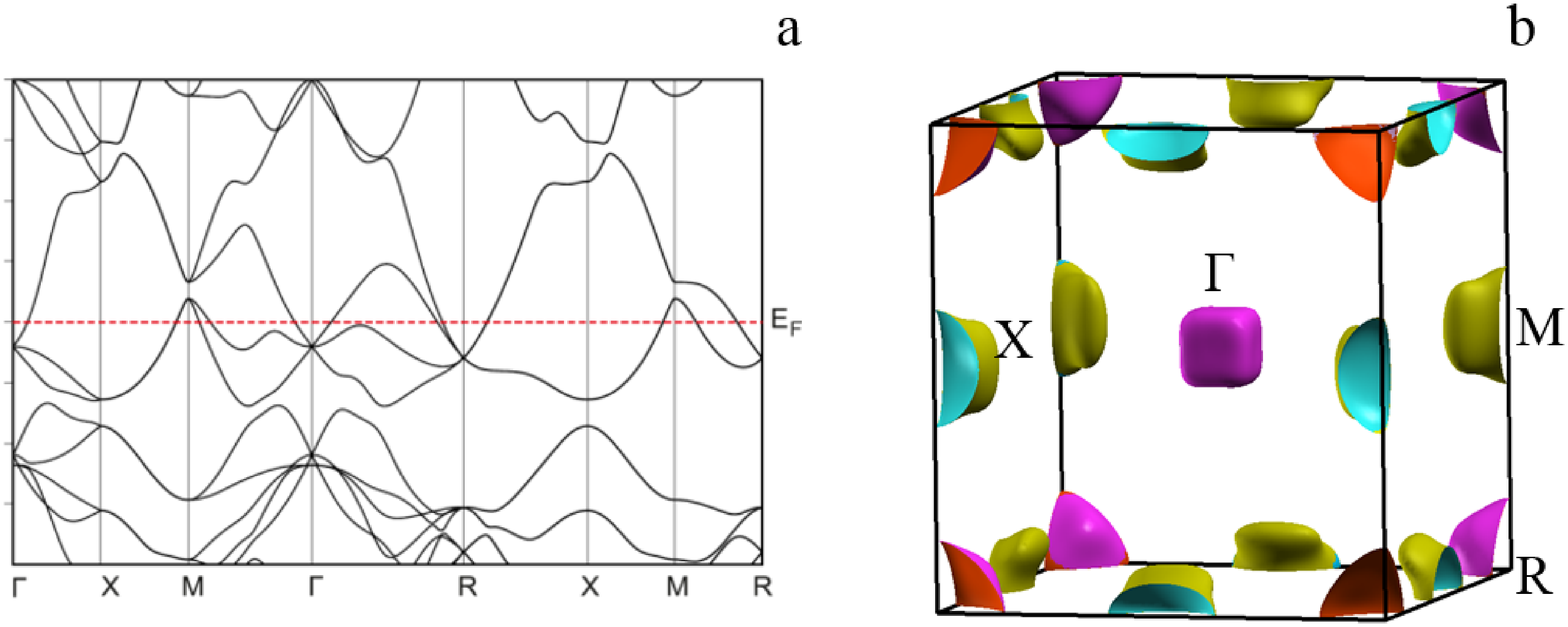}
\caption{\label{fig13}Paramagnetic RhGe: (a) The band structure along the high-symmetry lines. The red circle Energy is measured from the Fermi level, $E_{\mathrm{F}}$ (dashed line). (b) The Fermi surface with standard notation for the high-symmetry points of BZ.}
\end{figure}

\begin{figure*}
\includegraphics[width=\textwidth]{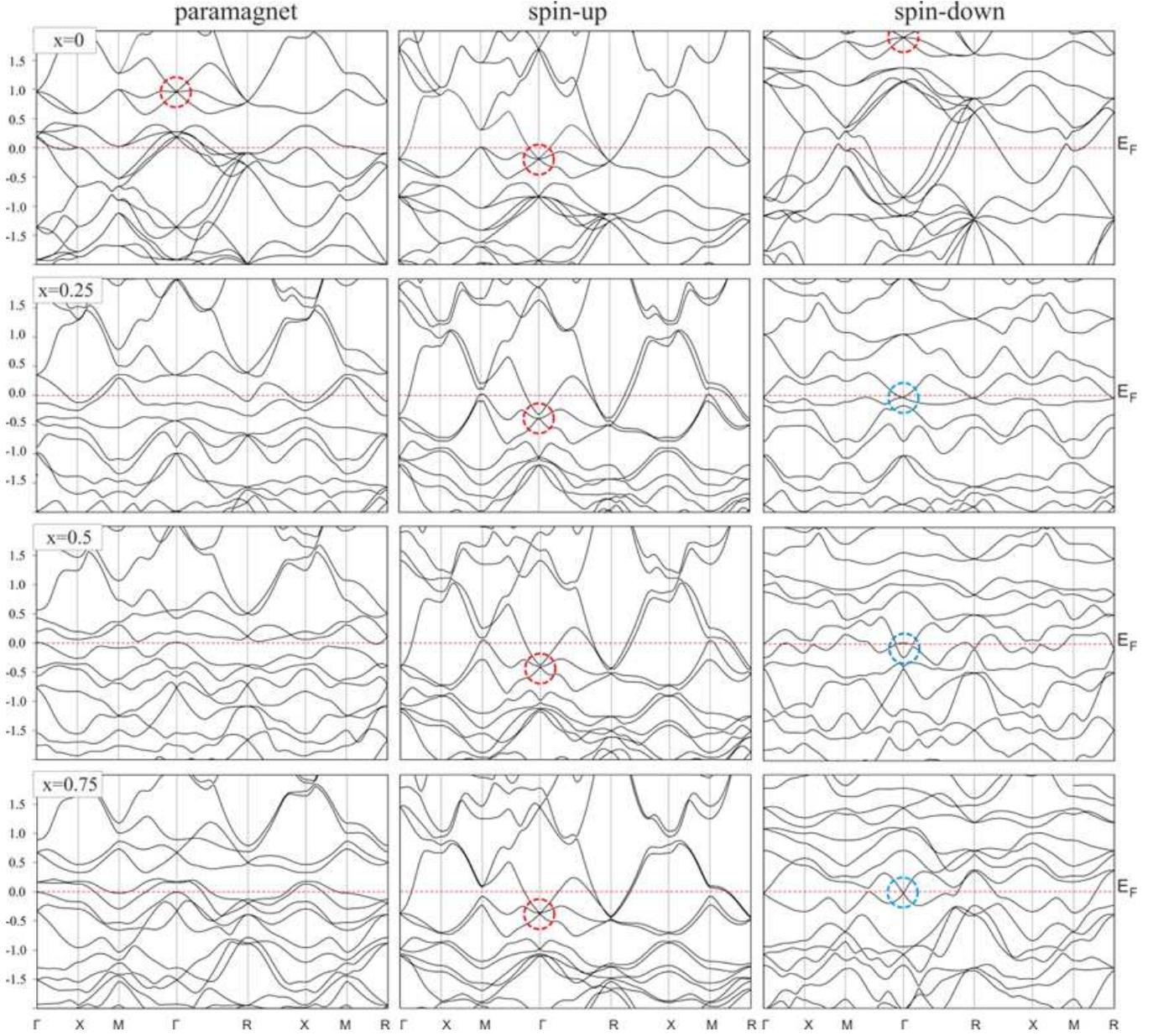}
\caption{\label{fig14}From top to bottom: Band structure along the high-symmetry lines at $x = 0, 0.25, 0.5$ and 0.75. Left-column plots relate to the paramagnetic state. The middle and right columns present, respectively, the spin-up and spin-down bands in the magnetic state. The Rh content $x$ is indicated in each row. All the spin-up band structures are rather similar to that of paramagnetic RhGe in FIG.~\ref{fig13}a. The red circles mark the symmetry-related points of triple degeneracy, blue circles, the 3D Dirac-like crossing points near $E_{\mathrm{F}}$ (see text). Note that the semiconductor-like band structure of Mn$_{0.5}$Rh$_{0.5}$Ge in the PM state becomes metal-like in the FM state. The energy is measured from the Fermi level, $E_{\mathrm{F}}$ (dashed line).}
\end{figure*}

Our calculated densities of states (DOS) of Mn$_{1-x}$Rh$_x$Ge in the PM and FM phase are presented in FIGs.~\ref{fig11} and \ref{fig12}a, correspondingly. FIGs.~\ref{fig14} and \ref{fig15} display the electronic band structure and Fermi surface (FS) at $x = 0, 0.25, 0.5$ and 0.75. Paramagnetic RhGe is shown separately in FIG.~\ref{fig13} as it represents a prototype for subsequent discussion. As is seen in the figures, the DOS and band structure of paramagnetic MnGe are very similar to those of RhGe. Actually, in both cases they are typical of nonmagnetic B20-type TM monogermanides \cite{23, 31, 32} and monosilicides \cite{33, 34}. This suggests that the electronic structure of B20 compounds is well described in the rigid band approximation, i.e. the position of the bands relative to the Fermi level, $E_{\mathrm{F}}$, is mainly determined by the electron count, while the general shape of bands remains practically unchanged. When a particular compound contains both the Mn and Rh atoms, the B20 lattice is distorted because of difference in the properties of $3d$- and $4d$-elements in their properties (atomic size, the space distribution of $d$ orbitals, the depth of potential well). Hence at $x = 0.25, 0.5$ and 0.75, the lattice can be only approximately considered as B20-type (our x-ray data also evidence that in the pseudobinary compounds studied, the B20 reflections are slightly broadened). It is clear that Mn$_{0.5}$Rh$_{0.5}$Ge possesses maximal lattice distortion among the compounds considered here.

FIG.~\ref{fig11} demonstrates that over the entire energy range, the DOS $N(E)$ of Mn$_{1-x}$Rh$_x$Ge is contributed mostly by transition-metal d-states hybridized with germanium $p$-states (not shown) with a dominating contribution from the former. At $x \leq 0.5$, the contribution from Mn $3d$-states exceeds that of Rh $4d$-states. The reason is that a more filled $d$-band of Rh lies deeper in energy (manganese has 5 $d$-electrons and rhodium 7; in MnGe and RhGe these values increase respectively, to 6.1 and 8.3 because of a charge transfer from $sp$- to $d$-band). Thus, the near-EF states in Mn$_{1-x}$Rh$_x$Ge, except for the Rh-rich side, are mainly formed by the electrons originating from the Mn atoms. A striking feature of the DOS in PM MnGe is a symmetry-conditioned gap situated at 0.5 eV above $E_{\mathrm{F}}$. As the Rh content $x$ increases, the gap is shifted towards lower energies and goes down to $E = -0.75$~eV in RhGe. At $x > 0.25$, it is actually a pseudogap, which is a consequence of increasing lattice distortion. At $x = 0.5$, the pseudogap falls exactly on the Fermi level (FIG.~\ref{fig11}c). It is clear if one bears in mind that the electron count of Mn$_{0.5}$Rh$_{0.5}$Ge coincides with that of FeGe which is insulating in the PM state \cite{35}.

The energy bands of the non-centrosymmetric and non-symmorphic B20 lattice have some symmetry-related peculiarities. For RhGe and MnGe in the PM state, all the bands along the R-X-M-R path (residing in the BZ face) are doubly degenerate and all the states at R point (the zone corner) are four-fold degenerate. Another uncommon feature is three intersecting bands at the zone center $\Gamma$ ($k = 0$), the middle of which has a zero slope (the electron group velocity $dE(k)/dk$), while the other two possess linear dispersions of equal slope and opposite sign, similar to Dirac cone This $\Gamma$-centered triply degenerate state lies just below $E_{\mathrm{F}}$ in RhGe and goes up to +1 eV in MnGe, owing to a smaller lattice parameter and a less filled $d$-band of the latter. At $0 < x < 1$, the degeneracy of bands is lifted by lattice distortion (FIG.~\ref{fig14}, left). Owing to the occurrence of zero-velocity bands below $E_{\mathrm{F}}$ (and related van Hove singularity), the Fermi level in RhGe falls on the verge of a wide valley of the DOS (FIG.~\ref{fig11}e). The Fermi surface of RhGe (FIG.~\ref{fig12}b) consists of $\Gamma$-centered electron-like pockets related to the three crossing bands, R-centered electron-like pockets, and hole-like pockets at the M point (middle of the zone edge). Notice that the FS of RhGe compressed to 8 GPa \cite{30} is analogous to that of isovalent nonmagnetic CoGe \cite{23}, which is a consequence of a smaller atomic size of Co.

The FS of MnGe in the PM state (FIG.~\ref{fig15}, left) consists of R-centered electron-like pockets and three large hole-like concentric voids on $\Gamma$ (shown transparent for clarity), the outer of which is open along the $\Gamma$-X direction (X denotes the center of BZ face). Inspection of the figure from top to bottom shows that the FS in the PM state dramatically changes with increasing $x$. At $x = 0.5$, the FS with its tiny hole- and electron-like pockets and a minimal charge-carrier density resembles that of semimetals. It only exists thanks to a very small indirect overlap of two bands near $E_{\mathrm{F}}$ (see FIG.~\ref{fig14}, left) corresponding to a pseudogap at $E_{\mathrm{F}}$. Upon further increase in $x$, a more metallic character of FS returns. In general, an increase in Rh content causes various void- and neck-type electronic topological transitions, which are expected to manifest themselves in anomalous concentration dependencies of electrical resistivity and thermopower.

According to our non-spin-polarized calculations, the least symmetric compound Mn$_{0.5}$Rh$_{0.5}$Ge in the PM state exhibits the worst metallic properties among the considered compositions. Qualitatively, this finding does not contradict to the semiconductor-like $T$-dependence of electrical resistivity observed in Mn$_{1-x}$Rh$_x$Ge above $T_m$, as $x$ lies in the interval $0.5 \pm 0.2$ (FIG.~\ref{fig4}). A qualitative correlation also exists between $x$-dependencies of the room-temperature Seebeck coefficient $S$ (FIG.~\ref{fig5}) and the DOS at the Fermi level $N(E_{\mathrm{F}})$ in the PM state (green curve in FIG.~\ref{fig12}b). The exception is represented by RhGe that possesses an anomalously large negative Seebeck coefficient that becomes large positive at a low hole doping when manganese substitutes for rhodium. An analogous behavior of $S(x)$ was observed for hole-doped B20-CoGe in Ref.~\cite{26}, where $S$ was calculated using the semi-classical Boltzmann approach within the relaxation-time approximation \cite{37}. It was demonstrated \cite{24} that the asymmetry of the density of states $N(E)$ related to the triply-degenerate bands just below $E_{\mathrm{F}}$ is a crucial factor to enhance $S$. In the near future we are about to quantitatively study the transport properties of Mn$_{1-x}$Rh$_x$Ge using appropriate software.

\subsection{Spin-polarized calculations}
FIG.~\ref{fig12}a displays the spin-majority (also denoted as `up` or `$\uparrow$`), spin-minority (`down` or `$\downarrow$`), and total (up+down) DOS at $x = 0, 0.25, 0.5$ and 0.75. Comparison between the concentration dependencies of the total DOS at the Fermi level, $N(E_{\mathrm{F}})$, in the PM and FM state is presented in FIG.~\ref{fig12}b. Contrary to the dramatic behavior in the PM state, the x-dependence of the total DOS at $E_{\mathrm{F}}$ in the FM state varies smoothly around $\approx 2.5$ eV$^{-1}$, a pronounced dip near $x = 0.5$ absent. The middle and right columns in FIG.~\ref{fig14} show the spin-split bands in the FM case. Let us consider first pure MnGe (upper row). Due to exchange splitting, the up and down subbands in FM MnGe are rigidly shifted respectively, downwards and upwards by $\approx 0.8$ eV with reference to the PM case, which yields a magnetization of about 2 $\mu_{\mathrm{B}}$/f.u. As a consequence of charge transfer to the low-lying majority bands, the resulting number of spin-up electrons $Z_\uparrow = 6.47$, a value very close to the valence of paramagnetic RhGe per spin, $Z = 6.5$ (FIG.~\ref{fig12}c). As is seen in the figures, the spin-up DOS, bands, and FS of MnGe are very similar in their shape to those of paramagnetic RhGe. The same is true for all the compounds with $0 < x < 1$.

FIG.~\ref{fig3} demonstrates that the total spin magnetic moment of Mn$_{1-x}$Rh$_x$Ge scales linearly with $x$ (or the valence $Z$). The calculated dependence $\mu(x)$ represents a negative-slope part (so called `itinerant`, usually related with fcc-derived structures) of the Slater-Pauling curve, where the magnetic moment linearly decreases with increasing $Z$. In this case, $E_{\mathrm{F}}$ is pinned within a valley of the spin-up DOS, as in RhGe. The valley constrains the number of the spin-up electrons, while the number of spin-down electrons increases proportionally to $Z$ and hence, so does the magnetic moment \cite{38}.

At all $x$, the spin-up bands not only retain stable positions with respect to $E_{\mathrm{F}}$, but also hold, to a large extent, a specific shape typical of B20 symmetry, while the spin-down bands at $0 < x < 1$ are strongly distorted (see FIG.~\ref{fig14}). Correspondingly, the triple degeneracy of levels is lifted, and instead, two-fold degenerate states (marked with blue circles) appear near $E_{\mathrm{F}}$. For example, the spin-down band structure of Mn$_{0.25}$Rh$_{0.75}$Ge is characterized by the intersection at $k = 0$ of two bands with close-to-linear dispersions $E({\mathbf{k}})$ of opposite sign (Dirac-like crossing point). These features deserve special consideration as a candidate 3D Dirac (or Weyl) points and will be the subject of future study.

\begin{figure*}
\includegraphics[width=\textwidth]{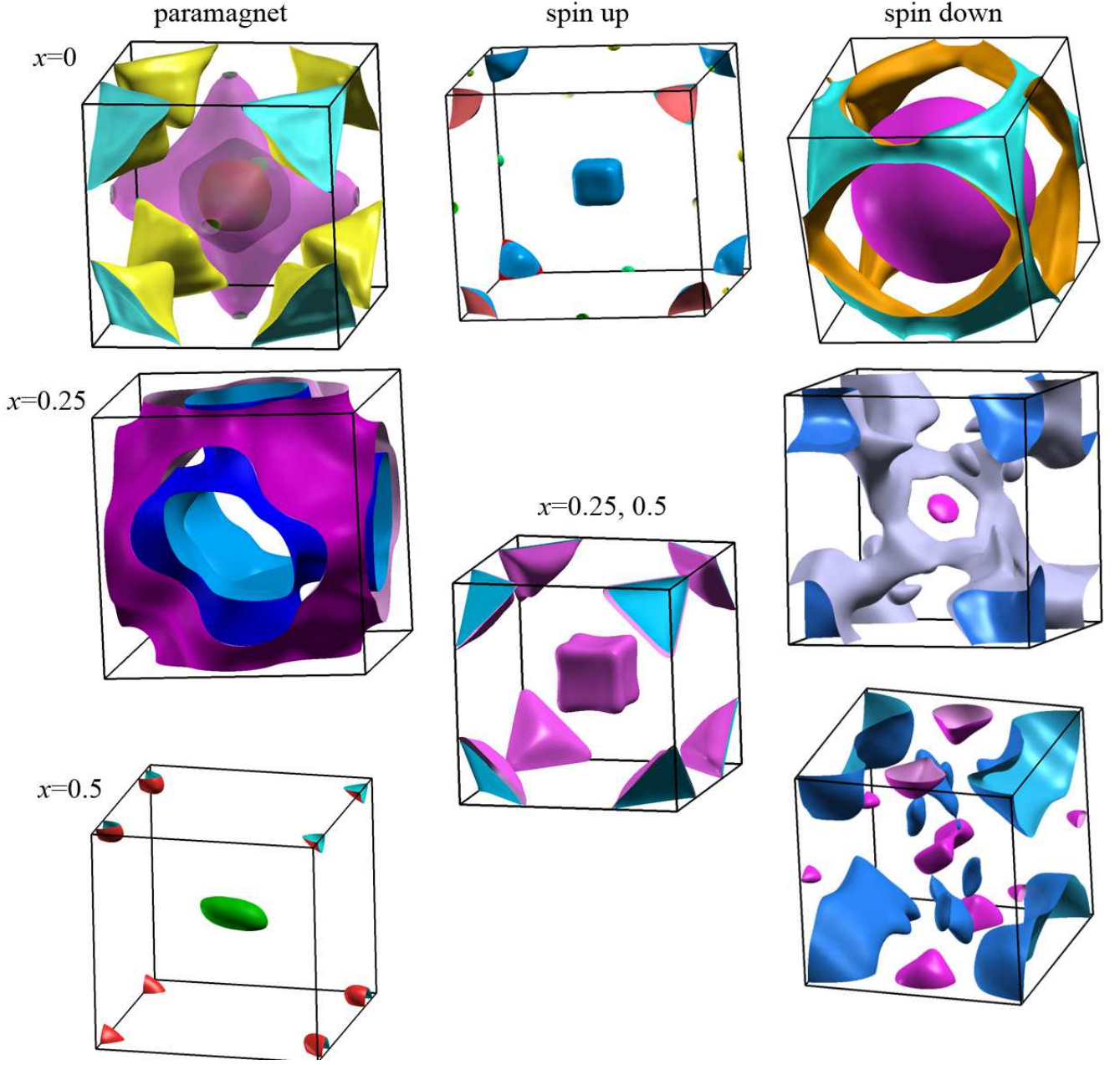}
\caption{\label{fig15}From top to bottom: Fermi surfaces at $x = 0, 0.25, 0.5$ and 0.75. Left-column plots relate to the paramagnetic state. The middle and right columns present, respectively, the spin-up and spin-down bands in the magnetic state. The Rh content $x$ is indicated in each row. The spin-up Fermi surfaces are topologically equivalent to that of paramagnetic RhGe (FIG.~\ref{fig13}b). The spin-up FSs at $x = 0.25$ and 0.5 are practically identical.}
\end{figure*}

The spin-resolved Fermi surfaces of magnetic Mn$_{1-x}$Rh$_x$Ge are presented in the middle and right columns of FIG.~\ref{fig15}. The spin-up FSs are topologically equivalent to that of paramagnetic RhGe in FIG.~\ref{fig13}b, with moderate variations in size of pockets around $\Gamma$, R and M points. At $x > 0.25$, the low-symmetry spin-down FSs consist of many irregular-shaped fragments, which reflects the growing lattice distortion. Again, we refer to paper \cite{23}, where the spin-up DOS, band structure, and FS of ferromagnetic MnGe and FeGe have been found similar to those of paramagnetic CoGe (isovalent to RhGe). To this we can add that the spin-down DOS, band structure, and FS of ferromagnetic FeGe coincide in every detail with those of paramagnetic MnGe calculated here and shown in FIGs.~\ref{fig11}, \ref{fig14} and \ref{fig15}.

As mentioned above, Mn$_{0.5}$Rh$_{0.5}$Ge in the PM state is a semimetal very close to an indirect-gap semiconductor. Our spin-polarized calculations demonstrate metallic character of the DOS, band structure and FS for all the considered compounds in the FM state, including Mn$_{0.5}$Rh$_{0.5}$Ge. The onset of metallicity in Mn$_{0.5}$Rh$_{0.5}$Ge at the transition to magnetic state is similar to the case of compressed isovalent FeGe, where collapse of magnetic moment has been found to coincide with a metal-to-semiconductor transition \cite{35}.

Even in case of disagreement between the measured and calculated lattice parameters, the magnitude of magnetic moment is known to be fairly well reproduced in standard DFT calculations. Here, the theoretical and experimental dependencies $\mu(x)$ well agree to each other (see FIG.~\ref{fig3}), a small systematic excess of calculated above measured values is most probably due to the approximation of collinear magnetism used in our calculation. The theoretical spin moment localized on Mn site, $\mu_{\mathrm{Mn}}$, very slightly increases with $x$, which corresponds to a lattice expansion due to increasing Rh content. As is seen in FIG.~\ref{fig3}, the cell magnetization is localized mostly at manganese atoms and virtually proportional to their number. The magnetic moments induced on the Rh and Ge sites (not shown in the figure) decrease with $x$. At all $x$, they are relatively small in magnitude ($\mu_{\mathrm{Rh}} < 0.05 \mu_{\mathrm{B}}$, $\mu_{\mathrm{Ge}} < 0.09 \mu_{\mathrm{B}}$) and parallel and antiparallel to $\mu_{\mathrm{Mn}}$, correspondingly. This result is consistent with the measured sign and $x$-dependence of the XMCD signal for Rh and Ge. In addition, we evaluated the magnetic moment on Mn site for Mn$_{0.5}$Rh$_{0.5}$Ge as a function of pressure $P$ and found that $\mu_{\mathrm{Mn}}(P)$ linearly decreases from 2.2 to 2.1 $\mu_{\mathrm{B}}$ upon compression to 6 GPa.

Our calculations reveal no stable magnetic solution for pure RhGe. Apparently, extremely small magnetization values ($\approx0.0007 \mu_{\mathrm{B}}$/f.u.) observed experimentally in RhGe \cite{13} are below the uncertainty of our calculations. However, within the itinerant magnetism model used here, the magnetization of RhGe is expected to be zero. Furthermore, according to available experimental and theoretical research, the isovalent B20 counterparts RhSi, CoSi, and CoGe are paramagnetic. We suppose that very weak ferromagnetism observed \cite{13} in high-pressure cubic phase of RhGe might be related to disordered moments or lattice defects and cannot be described within the simple band model. A more detailed study of magnetic properties of Mn$_{1-x}$Rh$_x$Ge, including non-collinearity is under way, the results will be published elsewhere. In this paper we undertook an initial theoretical investigation of the system Mn$_{1-x}$Rh$_x$Ge by means of standard DFT calculations, in an effort to reveal main trends in the concentration behavior of the system and find possible correlations with our experimental results.

\section{CONCLUSIONS}
To summarize, we experimentally studied the structural, magnetic and transport properties, as well as XANES/XMCD spectra, of the series of B20-type compounds Mn$_{1-x}$Rh$_x$Ge ($0 \leq x \leq 1$) synthesized under high pressure of 8 GPa. The system was also investigated theoretically on the basis of ab initio density-functional calculations. An anomalous non-linear concentration dependence of the lattice parameter, $a(x)$, was experimentally found for the high-pressure phases studied. This result is not explained with our DFT calculations, which produce a linear function $a(x)$, according to Vegard`s law. At the same time, the concentration dependencies of magnetization $\mu(x)$, both measured in external magnetic field and DFT-calculated, are in rather good agreement to each other and linearly decrease with increasing x. The total magnetization in unit cell is mostly localized on manganese atoms and virtually proportional to their number. The magnetic moment on Mn site ($\mu_{\mathrm{Mn}} \approx 2 \mu_{\mathrm{B}}$) is practically independent of concentration $x$. The significantly smaller magnetic moments on Rh and Ge sites decrease with $x$ and are directed parallel and antiparallel to $\mu_{\mathrm{Mn}}$, correspondingly. The comparison with our XMCD measurements of magnetic polarization induced on Rh and Ge atoms shows that our DFT calculations agree with observed sign and $x$-dependence of the XMCD signal for Rh and Ge.

Our measurements of magnetization and magnetic susceptibility reveal that as $x$ changes from zero to 0.3, the magnetic ordering temperature $T_m$ decreases from 170 K to 140 K. Upon further increase in rhodium concentration, $T_m$ rises again up to approximately 160 K at $x = 0.5$ and then slowly decreases down to $\approx 140$~K at $x = 0.975$. Under external compression up to 6 GPa, it was found that on the Mn-rich side (up to $x = 0.3$), $T_m$, as expected, decreases as a function of pressure P, while at $x \geq 0.5$, $T_m(P)$ definitely increases with elevated pressure (by 20\% at 6 GPa), the rate of the increase independent of $x$. Indirectly, this result is not confirmed theoretically, because our calculated magnetic moment linearly decreases with pressure. Such a pressure-induced increase in the magnetic ordering temperature is first observed for the B20-type high-pressure phases with helical magnetic structure, in contrast to all other $3d$-metal-based monosilicides and monogermanides with B20 structure. This unusual behavior needs a special theoretical justification and will be studied in the future.

The semiconductor-like behavior of electrical resistivity, $\rho(T)$, observed experimentally for paramagnetic Mn$_{1-x}$Rh$_x$Ge at $x$ lying in the interval $0.5 \pm 0.2$, is qualitatively consistent with our theoretical result that Mn$_{0.5}$Rh$_{0.5}$Ge in the PM state is very close to indirect-gap semiconductor, as judged from the shape of its electronic spectrum. Both our experiments and calculations at these intermediate concentrations demonstrate a semiconductor-to-metal transition, upon the occurrence of magnetic order. A qualitative correlation also exists between the concentration dependencies of the room-temperature Seebeck coefficient, $S(x)$, and the density of states at the Fermi level, $N(E_{\mathrm{F}})$, in the PM phase. An anomalously large negative Seebeck coefficient measured in RhGe is related to the triply-degenerate electron state, residing just below $E_{\mathrm{F}}$. Our $ab~initio$ calculations show that the electronic structure of Mn$_{1-x}$Rh$_x$Ge in the both PM and FM states is characterized by the presence near $E_{\mathrm{F}}$ of symmetry-conditioned points of three- and two-fold degeneracy, which deserve special consideration as candidate 3D Dirac (or Weyl) points and will be the subject of future study.

\begin{acknowledgments}
The authors gratefully thank S. M. Stishov for supporting of this work and I. Mirebeau, S. V. Grigoriev, Yu. A. Uspenskii, and A. V. Mikheenkov for valuable discussions. This work was supported by Russian Science Foundation: V.~A.~Sidorov, A.~E.~Petrova, D.~A.~Salamatin and A.~V.~Tsvyashchenko acknowledge the support of their experimental measurements (grant RSF 17-12-01050) and N. M. Chtchelkatchev is grateful for support of his theoretical calculations (grant RSF 18-12-00438). The support from Russian Foundation for Basic Research (grants 16-02-01122 and 17-02-00725) is also acknowledged. The numerical calculations were carried out using computing resources of the Federal Collective Usage Center Complex for Simulation and Data Processing for Mega-science Facilities at NRC `Kurchatov Institute`, http://ckp.nrcki.ru/. Part of calculations was performed at the cluster of Joint Supercomputing Center, Russian Academy of Sciences.
\end{acknowledgments}

\end{document}